\documentclass[12pt,aps,showpacs]{revtex4}%
\usepackage{amsmath}
\usepackage{amssymb}
\usepackage{amsfonts}
\usepackage{graphicx}%
\providecommand{\U}[1]{\protect\rule{.1in}{.1in}}
\begin{document}
\title{Velocity and\ Heat Flow in a Composite Two Fluid System }
\author{J.P. Krisch and E.N. Glass}
\affiliation{Department of Physics, University of Michigan,\ Ann Arbor, MI }
\date{14 November 2011}

\begin{abstract}
We describe the stress energy of a fluid with two unequal stresses and heat
flow in terms of two perfect fluid components. The description is in terms of
the fluid velocity overlap of the components, and makes no assumptions about
the equations of state of the perfect fluids. The description is applied to
the metrics of a conformally flat system and a black string.

\end{abstract}

\maketitle

\section{Introduction}

Descriptions of fluid systems used in general relativity have become more
complex as we acquire new data about our universe.\ Simple perfect fluids are
still useful models in many cases \cite{MV04}, but, increasingly, extensions
of the perfect fluid stress-energy to include anisotropy and fluid
interactions are necessary considerations.\ The stress-energy for an
anisotropic fluid can easily be written in terms of metric based tetrads
$[U_{a},R_{a},\Theta_{a},\Phi_{a}]$%
\begin{equation}
T_{ab}=(\varepsilon)\hat{U}_{a}\hat{U}_{b}+(P_{r})\hat{R}_{a}\hat{R}%
_{b}+(P_{\theta})\hat{\Theta}_{a}\hat{\Theta}_{b}+(P_{\phi})\hat{\Phi}_{a}%
\hat{\Phi}_{b}.
\end{equation}
without addressing the physical origins of the anisotropy. Fluid anisotropy
has many physical causes \cite{HS97,Iva10}. For a low density fluid,
anisotropy can be modeled by multi-perfect fluid descriptions with differences
in the component fluid velocities \cite{Jea22} or equations of state
\cite{BL74} generating the anisotropy. One method of incorporating these
differences is through stress-energy equivalence \cite{CW92,CT94} where a
composite system can be written as the sum of the two component perfect
fluids: \
\begin{equation}
(\varepsilon)\hat{U}_{a}\hat{U}_{b}+(P_{r})\hat{R}_{a}\hat{R}_{b}+(P_{\theta
})\hat{\Theta}_{a}\hat{\Theta}_{b}+(P_{\phi})\hat{\Phi}_{a}\hat{\Phi}%
_{b}=(\varepsilon_{1}+P_{1})\hat{U}_{a}^{(1)}\hat{U}_{b}^{(1)}+(\varepsilon
_{2}+P_{2})\hat{U}_{a}^{(2)}\hat{U}_{b}^{(2)}+(P_{1}+P_{2})g_{ab}
\label{2comp-1}%
\end{equation}
This form allows the component fluid motions and individual equations of state
to be built into the composite fluid description and has parameter freedom for
modeling assumptions. A related method is due to Letelier \cite{Let80}, where
he explicitly rotates the component fluid velocities to create a new composite
tetrad.\ Using the new tetrad, the composite stress-energy tensor describes a
fluid with one anisotropic stress, $\sigma, $ associated with a direction of
fluid anisotropy $\Upsilon_{a}.$
\begin{equation}
T_{ab}=(\varepsilon+\Pi)\hat{U}_{a}\hat{U}_{b}+(\sigma-\Pi)\hat{\Upsilon}%
_{a}\hat{\Upsilon}_{b}+(\Pi)g_{ab} \label{T-ij}%
\end{equation}
Letelier's stress-energy form follows from an assumption of zero heat flow in
the composite fluid and implies a relation between the stress-energies of the
component fluids, $\varepsilon_{1}+P_{1}=const\times(\varepsilon_{2}+P_{2}%
)$.\ Not all physical examples will obey this stress-energy relation or will
have zero heat flow, and Eq.(\ref{T-ij}) can follow from other physical
assumptions.\ Multi-fluid models of complex fluids are increasingly being
used, not only in formal general relativity \cite{CZ99} but also in other
physics sub areas. Two and three fluid descriptions cover phenomena like multi
charge species in magnetized plasmas \cite{PGB08}, superfluids \cite{TG05,
YMK07, THL+08}, cosmological models \cite{MSV09, MSV09b, SV10,APS11},
particles in heavy ion collisions \cite{IR07}, Fermi-Bose fluids
\cite{WTD02}\ and hydrodynamic nuclear models \cite{MDB71}.\ A need has
emerged for a range of two fluid descriptions to use as a basis for n-fluid
generalizations \cite{LA86,LGM04,Ull07,Wil11,MP11}. In this paper we suggest
an anisotropic two fluid stress form based on the overlap of the component
fluid velocity vectors rather than a stress-energy assumption.\ The
description allows heat flow along the direction of anisotropy in the
composite stress-energy and has no initial stress-energy assumptions. In the
next section we set up the stress-energy description.\ The relation between
the 4-velocity overlap and the stress-energy is discussed in the third
section, and some examples are discussed in the fourth part of the paper.

\section{Combining component fluids}

\subsection*{The stress-energy form}

Consider a manifold which contains two perfect fluid flows and metric $g_{ab}
$. A composite stress-energy for two perfect fluids is%
\begin{equation}
T_{ab}=(\varepsilon_{1}+P_{1})\hat{U}_{a}^{(1)}\hat{U}_{b}^{(1)}%
+(\varepsilon_{2}+P_{2})\hat{U}_{a}^{(2)}\hat{U}_{b}^{(2)}+g_{ab}(P_{1}%
+P_{2}). \label{composite-1}%
\end{equation}
In order to express the composite stress-energy in a simpler form, following
Letelier \cite{Let80}, we map the timelike unit vectors $[\hat{U}_{a}%
^{(1)},\hat{U}_{a}^{(2)}]$ to an un-normed pair $[U_{a}^{\ast},\Upsilon
_{a}^{\ast}]$ where $U_{a}^{\ast}$ is timelike and $\Upsilon_{a}^{\ast}$ is
spacelike. A general unimodular transformation between these two sets of
vectors can be written as%
\begin{equation}%
\begin{bmatrix}
\hat{U}_{a}^{(1)}\\
\hat{U}_{a}^{(2)}%
\end{bmatrix}
=%
\begin{bmatrix}
A\cos\alpha & -B\sin\alpha\\
D\sin\alpha & C\cos\alpha
\end{bmatrix}%
\begin{bmatrix}
U_{a}^{\ast}\\
\Upsilon_{a}^{\ast}%
\end{bmatrix}
\label{transf}%
\end{equation}
with unit transformation determinant $AC\cos^{2}\alpha+BD\sin^{2}\alpha=1 $
and transformation inverse
\begin{equation}%
\begin{bmatrix}
U_{a}^{\ast}\\
\Upsilon_{a}^{\ast}%
\end{bmatrix}
=%
\begin{bmatrix}
C\cos\alpha & B\sin\alpha\\
-D\sin\alpha & A\cos\alpha
\end{bmatrix}%
\begin{bmatrix}
\hat{U}_{a}^{(1)}\\
\hat{U}_{a}^{(2)}%
\end{bmatrix}
\label{inverse-transf}%
\end{equation}

Since $U^{\ast a}$ and $\Upsilon^{\ast a}$ are not normalized, two of the
constants can be absorbed into their definitions.\ Choosing $A=C=1$, the unit
determinant condition is
\begin{equation}
\cos^{2}\alpha+BD\sin^{2}\alpha=1 \label{bd-one}%
\end{equation}
The rotation angle $\alpha$ is fixed by requiring $g^{ab}\Upsilon_{a}^{\ast
}U_{b}^{\ast}=0$, with $U_{a}^{\ast}$ timelike, $\Upsilon_{a}^{\ast}$
spacelike and $g^{ab}\hat{U}_{a}^{(2)}\hat{U}_{b}^{(2)}=g^{ab}\hat{U}%
_{a}^{(1)}\hat{U}_{b}^{(1)}=-1$. The orthogonality condition implies
\begin{equation}
-(B-D)\cos\alpha\sin\alpha+\hat{U}^{(2)a}\hat{U}_{a}^{(1)}(\cos^{2}%
\alpha-BD\sin^{2}\alpha)=0, \label{overlap-eqn}%
\end{equation}
where $\hat{U}^{(2)a}\hat{U}_{a}^{(1)}$ is the velocity overlap. There are
several special values of $B$ which can be eliminated from the parameter range
by the requirement that $\Upsilon_{a}^{\ast}$ be spacelike. Condition $B=0$ in
determinant Eq.(\ref{bd-one}) implies $\alpha=0,\pi.$ From the transformation
equation Eq.(\ref{inverse-transf}), this choice also identifies both
$U_{a}^{\ast}$ and $\Upsilon_{a}^{\ast}$ as timelike vectors so that overlap
equation (\ref{overlap-eqn}) is not valid for values $B=0$, $\alpha=0,\pi
$.\ The choice $B=\pm1$ requires $D=\pm1$. The overlap equation thus becomes
$\hat{U}^{(2)a}\hat{U}_{a}^{(1)}\cos2\alpha=0,$ with $\alpha=\pi/4,3\pi/4.$
These values of $\alpha$ in cos$2\alpha=0$ are also excluded by the spacelike
condition on $\Upsilon_{a}^{\ast}$. From Eq.(\ref{inverse-transf}) with
$B=\pm1$, we have%
\begin{align}
\Upsilon_{a}^{\ast}  &  =\mp\sin\alpha\hat{U}_{a}^{(1)}+\cos\alpha\hat{U}%
_{a}^{(2)}\label{gamma-star}\\
U^{(1,2)}  &  :=\hat{U}_{a}^{(1)}\hat{U}^{(2)a}\label{overlap-def}\\
\Upsilon_{a}^{\ast}\Upsilon^{\ast a}  &  =-1\pm\sin2\alpha\lbrack-U^{(12)}]
\end{align}
The velocity overlap $U^{(1,2)}$ must be negative, since it is the product of
two future pointing timelike vectors. For $\Upsilon_{a}^{\ast}$ to be
spacelike requires $\pm\sin2\alpha\lbrack-U^{(12)}]>1,$ or $\alpha<\pi/4$ for
$B>0$. The ranges we will consider are $B\neq0,1$ and \ $0<\alpha<\pi
/4$.\ These ranges and the unit determinant condition require $BD=1$. The two
normalized unit vectors are%
\begin{equation}
\hat{U}^{b}=\frac{U^{\ast b}}{\sqrt{-U^{\ast a}U_{a}^{\ast}}},\text{ \ }%
\hat{\Upsilon}^{b}=\frac{\Upsilon^{\ast b}}{\sqrt{\Upsilon^{\ast a}%
\Upsilon_{a}^{\ast}}}, \label{unit-vec}%
\end{equation}
with norms
\begin{align}
\tilde{N}^{2}  &  =\Upsilon_{a}^{\ast}\Upsilon^{\ast a}=\frac{1}{2}\left[
\frac{1-B^{2}}{B^{2}\cos2\alpha}-(\frac{1+B^{2}}{B^{2}})\right] \\
N^{2}  &  =-U^{\ast a}U_{a}^{\ast}=\frac{1}{2}\left[  (\frac{1-B^{2}}%
{\cos2\alpha})+(1+B^{2})\right] \nonumber
\end{align}

\section{Overlap and Stress-Energy}

\subsection*{Stress-Energy}

The rotation angle can be expressed in terms of the 4-velocity overlap.%
\begin{align}
\tan2\alpha &  =-(\frac{2B}{1-B^{2}})U^{(1,2)}\\
B  &  \neq0,1\text{ \ \ \ }\ 0<\alpha<\pi/4\nonumber
\end{align}
The composite stress-energy tensor in Eq.(\ref{composite-1}) can now be
written as
\begin{equation}
T_{ab}=(\varepsilon+\Pi)\hat{U}_{a}\hat{U}_{b}+(\sigma-\Pi)\hat{\Upsilon}%
_{a}\hat{\Upsilon}_{b}+(\Pi)g_{ab}+Q_{a}\hat{U}_{b}+Q_{b}\hat{U}_{a}%
\end{equation}
with fluid parameters
\begin{subequations}
\begin{align}
\Pi &  =p_{1}+p_{2}\\
Q_{a}  &  =\hat{\Upsilon}_{a}N\tilde{N}[B^{-1}(\varepsilon_{2}+P_{2}%
)-B(\varepsilon_{1}+P_{1})]\sin\alpha\cos\alpha\\
\varepsilon+\Pi &  =N^{2}[(\varepsilon_{1}+P_{1})\cos^{2}\alpha+B^{-2}%
(\varepsilon_{2}+P_{2})\sin^{2}\alpha]\\
\sigma-\Pi &  =\tilde{N}^{2}[B^{2}(\varepsilon_{1}+P_{1})\sin^{2}%
\alpha+(\varepsilon_{2}+P_{2})\cos^{2}\alpha]
\end{align}
and $B$ an unspecified constant except for $B\neq0,1$. \ 

\subsection*{Heat Flow and B}

Letelier's \cite{Let80} choice, $B^{2}=(\varepsilon_{2}+p_{2})/(\varepsilon
_{1}+p_{1})$\ reproduces the original two fluid stress-energy with $Q_{a}%
=0$.\ However, the underlying physics of the description can depend on both
the velocity overlap and general equations of state in the component
fluids.\ For example, if the two fluids move together with the same
4-velocity, the overlap is $-1$ and we have tan$(2\alpha_{0})=2B/(1-B^{2}).$
Another way of choosing $B$ that allows non zero heat flow is to use the
$\hat{U}_{a}^{(1)}=\hat{U}_{a}^{(2)}$ condition to define $B$ as
\end{subequations}
\begin{equation}
B=\tan\alpha_{0}%
\end{equation}
with $\alpha=\alpha_{0}$ producing the 'aligned' fluid.\ Substituting the
alignment condition in Eq.(\ref{gamma-star}) requires the aligned fluid to
have zero anisotropy vector, $\Upsilon_{a}^{\ast}=0$ \ [recall, $U_{a}^{\ast
}=\cos\alpha\hat{U}_{a}^{(1)}+B\sin\alpha\hat{U}_{a}^{(2)}$, $\ \Upsilon
_{a}^{\ast}=(-1/B)\sin\alpha\hat{U}_{a}^{(1)}+\cos\alpha\hat{U}_{a}^{(2)}%
$].\ The conditions on $B$ set the positive range $0<\alpha_{0}<\pi/4,$ such
that $\alpha_{0}\leq\alpha<\pi/4.$\ \ The velocity overlap is%
\begin{equation}
\tan2\alpha=-U^{(1,2)}\tan2\alpha_{0}%
\end{equation}
The\ general normalizations are
\begin{align}
\tilde{N}^{2}  &  =\Upsilon_{a}^{\ast}\Upsilon^{\ast a}=\frac{1}{2\sin
^{2}\alpha_{0}}\left[  \frac{\cos2\alpha_{0}}{\cos2\alpha}-1\right] \\
N^{2}  &  =-U^{\ast a}U_{a}^{\ast}=\frac{1}{2\cos^{2}\alpha_{0}}\left[
\frac{\cos2\alpha_{0}}{\cos2\alpha}+1\right]
\end{align}
Note that both $\tilde{N}$ and $\Upsilon_{a}^{\ast}$ have zero values when the
fluid is aligned with $\alpha=\alpha_{0}$. The fluid parameters are
\begin{subequations}
\label{fluid-param}%
\begin{align}
\Pi &  =P_{1}+P_{2}\\
\varepsilon+\Pi &  =\frac{2}{\sin^{2}2\alpha_{0}}\left[  \frac{\cos2\alpha
_{0}}{\cos2\alpha}+1\right]  [(\varepsilon_{1}+P_{1})\sin^{2}\alpha_{0}%
\cos^{2}\alpha+(\varepsilon_{2}+P_{2})\cos^{2}\alpha_{0}\sin^{2}\alpha]\\
\sigma-\Pi &  =\frac{2}{\sin^{2}2\alpha_{0}}\left[  \frac{\cos2\alpha_{0}%
}{\cos2\alpha}-1\right]  [(\varepsilon_{1}+P_{1})\sin^{2}\alpha_{0}\sin
^{2}\alpha+(\varepsilon_{2}+P_{2})\cos^{2}\alpha_{0}\cos^{2}\alpha]
\end{align}

The heat flow vector is
\end{subequations}
\begin{equation}
Q_{a}=\left(  \frac{\sin\alpha\cos\alpha}{2\sin^{2}\alpha_{0}}\sqrt
{(\frac{\cos2\alpha_{0}}{\cos2\alpha})^{2}-1}\ [(\varepsilon_{2}+P_{2}%
)-\tan^{2}\alpha_{0}(\varepsilon_{1}+P_{1})]\right)  \hat{\Upsilon}_{a}%
\end{equation}
and is zero for $\alpha=\alpha_{0}.$

\subsection*{The $\alpha,\alpha_{0}$ boundary}

The $\hat{U}_{a}^{(1)}=\hat{U}_{a}^{(2)}$ alignment condition has
stress-energy relations $Q=0$ and $\varepsilon+\Pi=\varepsilon_{1}%
+P_{1}+\varepsilon_{2}+P_{2}$. When $\alpha$ is close to $\alpha_{0},$ the
4-velocities should be only slightly different and the fluids should show
small deviations from the aligned relations. There are three parameter
boundary cases: (1) both $\alpha$ and $\alpha_{0}$ near zero, (2) both
$\alpha$ and $\alpha_{0}$ near $\pi/4,$ and (3) $\alpha_{0}$ near zero and
$\alpha$ just under $\pi/4$.

One expects the first two cases to describe a composite fluid only slightly
different from the aligned composite.\ For the first case,\ $\alpha_{0}%
=\delta_{0},$ $\alpha=\delta$, choose $\alpha_{0}$ and $\alpha$ small but of
the same order with $\delta_{0}<\delta.$\ We have $R\approx1+2\delta
^{2}-2\delta_{0}^{2}$ and $\tan2\alpha/\sqrt{R^{2}-1}\approx\delta
/\sqrt{\delta^{2}-\delta_{0}^{2}}>>1.$ The fluid parameters are
\begin{subequations}
\begin{align}
-U^{(12)}  &  \sim\delta/\delta_{0}\\
\varepsilon+\Pi &  \sim(\varepsilon_{1}+P_{1})+(\varepsilon_{2}+P_{2}%
)(\delta/\delta_{0})^{2}\\
\sigma-\Pi &  \sim[(\delta/\delta_{0})^{2}-1](\varepsilon_{2}+P_{2})
\end{align}
with heat flow
\end{subequations}
\[
Q\sim(\delta/\delta_{0})\sqrt{(\delta/\delta_{0})^{2}-1}\ (\varepsilon
_{2}+P_{2})
\]
The second case, $\alpha=\pi/4-\delta,$ $\alpha_{0}=\pi/4-\delta_{0}$, is very
similar to the first with $\delta_{0}>\delta.$ The velocity overlap is again,
almost aligned, $-U^{(1,2)}\sim\delta_{0}/\delta$, but the composite fluid
relations to the component fluids are multiples of the aligned fluid
description. \ \
\begin{subequations}
\begin{align}
Q  &  \sim\frac{1}{2}\sqrt{(\delta_{0}/\delta)^{2}-1}\ [(\varepsilon_{2}%
+P_{2}-(\varepsilon_{1}+P_{1})]\\
\varepsilon+\Pi &  \sim\frac{1}{2}\left[  \delta_{0}/\delta+1\right]
[\varepsilon_{1}+P_{1}+\varepsilon_{2}+P_{2}]\\
\sigma-\Pi &  \sim\frac{1}{2}\left[  \delta_{0}/\delta-1\right]
[\varepsilon_{1}+P_{1}+\varepsilon_{2}+P_{2}]
\end{align}
The stress-energy in both these cases obeys the $U_{a}^{(1)}=U_{a}^{(2)}$
condition to lowest order, $\varepsilon-\sigma+2\Pi\sim\varepsilon_{1}%
+P_{1}+\varepsilon_{2}+P_{2}.$ The physical difference between the two cases
can be explained by the composite $\ast$ vector relation to the component
fluid velocities,\ Eq.(\ref{inverse-transf}). For parameter values near zero,
the component fluid velocity is dominated by the first fluid, $U_{a}^{\ast
}\sim\hat{U}_{a}^{(1)}$\ with direction of anisotropy $\Upsilon_{a}^{\ast}%
\sim-(\delta/\delta_{0})\hat{U}_{a}^{(1)}+\hat{U}_{a}^{(2)}$. For parameter
values near $\pi/4,\ U_{a}^{\ast}\sim(\hat{U}_{a}^{(1)}+\hat{U}_{a}%
^{(2)})/\sqrt{2}\ $and $\Upsilon_{a}^{\ast}\sim(-\hat{U}_{a}^{(1)}+\hat{U}%
_{a}^{(2)})/\sqrt{2}$. The straight combination of velocities in the second
case explaining its close similarity to the aligned fluid case. The third case
is an example of strong non-alignment. When $\alpha$ and $\alpha_{0}$ are at
opposite ends of the parameter range, $\alpha_{0}=\delta_{0},\ \alpha
=\pi/4-\delta$, we have $R=\cos2\delta_{0}/\sin2\delta>>1$ and $\tan
(\pi/2-2\delta)/\sqrt{R^{2}-1}\approx1.$ The fluid parameters obey a very
different relation than the aligned fluid with $\varepsilon+2\Pi-\sigma\sim0$,
rather than the sum of the component stress-energy. We also have \ \
\end{subequations}
\begin{subequations}
\begin{align}
\varepsilon+\Pi &  \sim\frac{(\varepsilon_{2}+P_{2})}{8\delta\delta_{0}^{2}}\\
\sigma-\Pi &  \sim\frac{(\varepsilon_{2}+P_{2})}{8\delta\delta_{0}^{2}}\\
Q  &  \sim\frac{(\varepsilon_{2}+P_{2})}{8\delta\delta_{0}^{2}}%
\end{align}
For this case the $\ast$ vectors are related to the component velocities by
$U_{a}^{\ast}\sim(\hat{U}_{a}^{(1)}+\delta_{0}\hat{U}_{a}^{(2)})/\sqrt{2}$ and
$\Upsilon_{a}^{\ast}\sim(-\hat{U}_{a}^{(1)}/\delta_{0}+\hat{U}_{a}%
^{(2)})/\sqrt{2}$. In the next section we give some metric examples. \ 

\section{Applications}

As an application of the two perfect fluid description we consider three
different examples.\ The first two are metric based with an anisotropic
stress-energy following from the field equations.\ The inverse of
Eqs.(\ref{fluid-param}b,\ref{fluid-param}c) give the component stress-energies
in terms of the composite descriptions: \
\end{subequations}
\begin{align}
\varepsilon_{1}+P_{1} &  =\cos^{2}\alpha_{0}\left[  (\varepsilon+\Pi
)\frac{\cos2\alpha+1}{\cos2\alpha_{0}+\cos2\alpha}+(\sigma-\Pi)\frac
{\cos2\alpha-1}{\cos2\alpha_{0}-\cos2\alpha}\right]  \\
\varepsilon_{2}+P_{2} &  =\sin^{2}\alpha_{0}\left[  (\varepsilon+\Pi
)\frac{\cos2\alpha-1}{\cos2\alpha_{0}+\cos2\alpha}+(\sigma-\Pi)\frac
{\cos2\alpha+1}{\cos2\alpha_{0}-\cos2\alpha}\right]
\end{align}
The two metric examples are a conformally flat spacetime and a black
string.\ The stress-energy from the field equations for both spacetimes has
$\varepsilon+\Pi=0$,$\ $and we can describe the related component perfect
fluids. The third example uses two dusty component perfect fluids\ and a
single anisotropic stress with no heat flow. The component fluid density
relations are examined along with the equation of state in the composite.\ For
this example, the Letelier description and the description in this paper coincide.

\subsection*{Example: A conformally flat spacetime}

A simple conformally flat spacetime has metric and field generated fluid
parameters as seen by a comoving observer $\hat{U}^{a}=(e^{-az},0,0,0).$
\begin{subequations}
\begin{align}
ds^{2}  &  =e^{2az}(-dt^{2}+dr^{2}+r^{2}d\phi^{2}+dz^{2}),\\
8\pi\varepsilon &  =-a^{2}e^{-2az},\\
8\pi\Pi &  =a^{2}e^{-2az},\\
\Pi &  =P_{r}=P_{\phi},\\
8\pi\sigma &  =8\pi P_{z}=3a^{2}e^{-2az},
\end{align}
with the anisotropy in the z-direction, $\Upsilon^{a}=(0,0,0,e^{-az})$. The
z-dependent negative density in this solution does not lend itself to a
physical description outside of a cosmological constant or Casimir effects.
The two fluid description of the composite fluid model has the advantage of
explaining the negative composite density in terms of tension in the first
fluid. For this fluid $\varepsilon+\Pi=0$ and $8\pi(\sigma-\Pi)=2a^{2}%
e^{-2az}$. From Eq.(\ref{fluid-param}), the component fluid parameters that
could create a composite fluid are
\end{subequations}
\begin{align}
\varepsilon_{1}+P_{1}  &  =\frac{2a^{2}e^{-2az}}{8\pi}\frac{(\cos
2\alpha-1)\cos^{2}\alpha_{0}}{(\cos2\alpha_{0}-\cos2\alpha)}\\
\varepsilon_{2}+P_{2}  &  =\frac{2a^{2}e^{-2az}}{8\pi}\frac{(\cos
2\alpha+1)\sin^{2}\alpha_{0}}{(\cos2\alpha_{0}-\cos2\alpha)}%
\end{align}
The second fluid could have both positive stress and density.\ The first
fluid, if it has positive density, must describe a fluid with tension rather
than pressure.\ The heat flow is axial%
\begin{equation}
Q_{a}=\left(  \frac{a^{2}e^{-2az}\sin2\alpha}{8\pi}\ \sqrt{\frac{\cos
2\alpha_{0}+\cos2\alpha}{\cos2\alpha_{0}-\cos2\alpha}}\right)  \ \hat
{\Upsilon}_{a}%
\end{equation}
The 4-velocities of the component fluid come directly from the transformation
equations and involve, unsurprisingly, a coordinate velocity in the
z-direction. \
\begin{align}
U^{(1)a}  &  =[e^{-az}N\cos\alpha,\text{ }0,\text{ }0,\text{ }-e^{-az}%
\tilde{N}\tan\alpha_{0}\sin\alpha]\label{conf-vel1}\\
U^{(2)a}  &  =[e^{-az}N\frac{\sin\alpha}{\tan\alpha_{0}},\text{ }0,\text{
}0,\text{ }e^{-az}\tilde{N}\cos\alpha] \label{conf-vel2}%
\end{align}
$\hat{U}^{(i)a}\hat{U}_{a}^{(i)}=-1$ was imposed in Eq.(\ref{unit-vec}). Both
$U^{(1)a}$ and $U^{(2)a}$ are expanding, accelerating and
shear-free.\ Requiring the velocity overlap, Eq.(\ref{overlap-def}) to be
constant, $U^{(12)}=const$, can be restrictive for shear-free fluids,
requiring acceleration and expansion (see Appendix A).\ That restriction is
met for this example.

\subsection*{Example: A black\ string}

A similar example to the conformal metric is the\ black string of Lemos and
Zanchin \cite{LZ96}, with metric%
\begin{equation}
ds^{2}=-(\Lambda r^{2}-m)dt^{2}+\frac{dr^{2}}{(\Lambda r^{2}-m)}+r^{2}%
d\phi^{2}+dz^{2}%
\end{equation}
and comoving stress-energy following from the field equations
\begin{subequations}
\begin{align}
8\pi\varepsilon &  =-\Lambda,\\
8\pi\Pi &  =\Lambda,\\
8\pi\sigma &  =8\pi P_{z}=3\Lambda.
\end{align}
From the first example, the replacement $a^{2}e^{-2az}$ $\rightarrow\Lambda$,
gives the component stress-energy and heat flow for this case. The direction
of anisotropy is along the string axis.\ Here, since the cosmological constant
can be negative and does not depend on position, a negative energy density is
possible. The fluid matter obeys $\varepsilon+\Pi=0$ and $\sigma=3\Pi\ $. The
4-velocities for the black string components are
\end{subequations}
\begin{align}
U^{(1)a}  &  =[\frac{N}{\sqrt{\Lambda r^{2}-m}}\cos\alpha,\text{ }0,\text{
}0,\text{ }-\tilde{N}\tan\alpha_{0}\sin\alpha]\\
U^{(2)a}  &  =[\frac{N}{\sqrt{\Lambda r^{2}-m}}\frac{\sin\alpha}{\tan
\alpha_{0}},\text{ }0,\text{ }0,\text{ }\tilde{N}\cos\alpha]
\end{align}
These component velocities are expansion-free but have acceleration, shear,
and vorticity. \ 

\subsection*{Example: Linear composite equation of state}

For the third example consider a composite fluid with $\Pi=0$, and two
component dust fluids, $\ P_{1}=P_{2}=0$. The composite fluid has a linear
equation of state related to $\alpha$ and $\alpha_{0}$:
\begin{equation}
\frac{\varepsilon}{\frac{\cos2\alpha_{0}}{\cos2\alpha}+1}=\frac{\sigma}%
{\frac{\cos2\alpha_{0}}{\cos2\alpha}-1}%
\end{equation}
With Eq.(\ref{fluid-param}) we have for the composite fluid
\begin{equation}
\varepsilon_{1}\sin^{2}\alpha_{0}=\varepsilon_{2}\cos^{2}\alpha_{0}%
\end{equation}
and this example has the Letelier \cite{Let80} stress-energy relation with
zero heat flow.
\begin{align}
\varepsilon &  =\frac{\varepsilon_{1}}{2\cos^{2}\alpha_{0}}\left[  \frac
{\cos2\alpha_{0}}{\cos2\alpha}+1\right] \\
\sigma &  =\frac{\varepsilon_{1}}{2\cos^{2}\alpha_{0}}\left[  \frac
{\cos2\alpha_{0}}{\cos2\alpha}-1\right]
\end{align}
The composite fluid has a dusty equation of state for $\alpha$ close to
$\alpha_{0}$, $R=(\cos2\alpha_{0}/\cos2\alpha)\sim1$, changing to a stiff
equation of state for large $R$.

\section{DISCUSSION}

In conclusion, we have presented an anisotropic fluid recipe based on
component velocity overlap. It is useful in describing anisotropic fluids in
terms of possible perfect fluid components, and may also be of use in
developing broader descriptions of multi-fluid models.\ The description
includes heat flow driven by fluid velocity non-alignment.\ The heat flow is
along the direction of anisotropy in the stress-energy form.\ Heat flows with
several spatial vector components are possible \cite{CZ99}, especially if
spatial tetrads are chosen to describe acceleration, or a null-vector
construction, or other non-heat related parameters.\ Including more general
heat descriptions is an idea for future work. The fluids considered here are
non-interacting; another possible extension is to component fluids with non
zero, but balancing stress-energy divergence \cite{KM08, CMM+08, P-NF08,
BC06}, and to acoustic phenomena related to the heat flow
\cite{Cis01,ZQ04,Kos11}.

\appendix{}

\section{Restrictions on the Component Fluids}

The behavior of the composite fluid is determined by the component velocity
overlap and equation of state. The velocity overlap equation following from
tetrad orthogonality can be restrictive since it requires the velocity overlap
to be constant, $U^{(1,2)}=const$. \ This will impose restrictions on the
component fluids. Consider the covariant derivative of the constant velocity
overlap $\nabla_{b}U^{(1,2)}=0$
\begin{equation}
\hat{U}^{(1)a}U_{a;b}^{(2)}=-U_{;b}^{(1)a}\hat{U}_{a}^{(2)}%
\end{equation}
Each of the velocity derivatives can be expanded in terms of its acceleration,
expansion, shear and vorticity ($\kappa^{i},\Theta^{i},$ $\sigma_{ab}^{i},$
$\omega_{ab}^{i})$ and projection operator $h_{ab}^{(i)}=g_{ab}+\hat{U}%
_{a}^{(i)}\hat{U}_{b}^{(i)}.$ \
\begin{equation}
\lbrack-\hat{U}_{a}^{(1)}\hat{U}_{b}^{(1)}+\sigma_{1ab}+\omega_{1ab}%
+\frac{\theta_{1}}{3}h_{ab}^{(1)}]\hat{U}^{(2)a}=-\hat{U}^{(1)a}[-\dot{U}%
_{a}^{(2)}\hat{U}_{b}^{(2)}+\sigma_{2ab}+\omega_{2ab}+\frac{\theta_{2}}%
{3}h_{ab}^{(2)}]
\end{equation}
The fluid acceleration can be parameterized with the Frenet tetrad associated
with each velocity vector\ $[\hat{U}_{a}^{(i)},$ $\hat{A}_{a}^{(i)},$ $\hat
{B}_{a}^{(i)},$ $\hat{C}_{a}^{(i)}]$.\ The acceleration lies along the vector
$\hat{A}_{a}^{(i)}$ and can be written as $\dot{U}_{a}^{(i)}=U_{a;b}%
^{(i)}U^{(i)b}=\kappa A_{a}^{(i)}.$ The acceleration for fluid one can be
isolated with $U^{(1)b}$ multiplication and similarly, for fluid two with
$U^{(2)b}$ multiplication:%
\begin{align*}
\kappa_{1}\hat{A}_{a}^{(1)}\hat{U}^{(2)a}  &  =\kappa_{2}\hat{A}_{a}^{(2)}%
\hat{U}^{(1)a}U^{(1,2)}-\sigma_{2ab}\hat{U}^{(1)a}\hat{U}^{(1)b}-\frac
{\theta_{2}}{3}[-1+(U^{(1,2)})^{2}]\\
\kappa_{2}\hat{A}_{a}^{(2)}\hat{U}^{(1)a}  &  =\kappa_{1}\hat{A}_{a}^{(1)}%
\hat{U}^{(2)a}U^{(1,2)}-\sigma_{1ab}\hat{U}^{(2)a}\hat{U}^{(2)b}-\frac
{\theta_{1}}{3}[-1+(U^{(1,2)})^{2}]
\end{align*}
If the component fluids are shear-free we have%
\begin{equation}
\kappa_{1}\hat{A}_{a}^{(1)}\hat{U}^{(2)a}=\kappa_{2}\hat{A}_{a}^{(2)}\hat
{U}^{(1)a}U^{(1,2)}-\frac{\theta_{2}}{3}[-1+(U^{(1,2)})^{2}]
\end{equation}%
\begin{equation}
\kappa_{2}\hat{A}_{a}^{(2)}\hat{U}^{(1)a}=\kappa_{1}\hat{A}_{a}^{(1)}\hat
{U}^{(2)a}U^{(1,2)}-\frac{\theta_{1}}{3}[-1+(U^{(1,2)})^{2}]
\end{equation}
If the two component fluids are shear free and have no expansion then they
satisfy%
\begin{align*}
\kappa_{1}A_{a}^{(1)}U^{(2)a}  &  =\kappa_{1}A_{a}^{(1)}U^{(2)a}%
(U^{(1,2)})^{2}\\
\kappa_{2}A_{a}^{(2)}U^{(1)a}  &  =\kappa_{2}A_{a}^{(2)}U^{(1)a}%
(U^{(1,2)})^{2}%
\end{align*}
If the expansions are zero, this can only be satisfied by identical component
fluid velocities, $U^{(1,2)}=-1$ or by zero accelerations, $\kappa_{1}%
,\kappa_{2}=0.$ The component fluids must be unaccelerated or aligned. If the
fluids are not aligned or if there is acceleration, one (or both) must have
expansion if they are shear-free.\ In the first metric example, the component
velocities\ Eqs.(\ref{conf-vel1},\ref{conf-vel2}) are shear and vorticity
free, and have both expansion and acceleration.\ The black string component
fluids have zero expansion but non-zero radial acceleration, with shear
components $\sigma_{0r}$ and $\sigma_{zr}$, and non-zero vorticity. Collins
\cite{Col86} has conjectured that, if shear-free perfect fluids obey a
barotopic equation of state, then either the expansion or vorticity must be
zero. If the conjecture is true \cite{Sop98, VdB99,VCK07,CKV+09}, the
shear-free, accelerated, unaligned component fluids must have zero vorticity
or a non-barotopic equation of state. The conjecture holds for the conformal
metric and does not apply to the black string. \

\end{document}